\documentclass[sn-standardnature,pdflatex]{sn-jnl}% Standard Nature Portfolio Reference Style

\usepackage{lipsum,amsmath,amssymb,bm}
\usepackage{graphicx,color}
\usepackage[normalem]{ulem}
\pdfoutput=1

\jyear{2022}

\begin{document}

\title[Defect dynamics in active smectics]{Defect dynamics in active smectics induced by confining geometry and topology}

\author*[1]{Zhi-Feng Huang}\email{huang@wayne.edu}
\author*[2]{Hartmut L\"owen}\email{hartmut.loewen@uni-duesseldorf.de}
\author*[3]{Axel Voigt}\email{axel.voigt@tu-dresden.de}

\affil[1]{Department of Physics and Astronomy, Wayne State University, Detroit, MI 48201, USA}
\affil[2]{Institut f\"ur Theoretische Physik II: Soft Matter, Heinrich-Heine-Universit\"at D\"usseldorf, 40225 D\"usseldorf, Germany}
\affil[3]{Institute of Scientific Computing, Technische Universit\"at Dresden, 01062 Dresden, Germany}

\abstract{
The persistent dynamics in systems out of equilibrium, particularly those characterized by annihilation and creation of topological defects, is known to involve complicated spatiotemporal processes and is deemed difficult to control. Here the complex dynamics of defects in active smectic layers exposed to strong confinements is explored, through self-propulsion of active particles and a variety of confining geometries with different topology, ranging from circular, flower-shaped epicycloid, to hypocycloid cavities, channels, and rings. We identify a wealth of dynamical behaviors during the evolution of complex spatiotemporal defect patterns as induced by the confining shape and topology, particularly a perpetual creation-annihilation dynamical state at intermediate activity with large fluctuations of topological defects and a controllable transition from oscillatory to damped time correlation of defect number density via mechanisms governed by boundary cusps. Our results are obtained by using an active phase field crystal approach. Possible experimental realizations are also discussed.
}

\maketitle

\section*{Introduction}

Defects in ordered or pattern-forming systems are of great interest both from a fundamental physics point of view highlighting the role of topology in condensed matter \cite{chaikin_lubensky_1995all,PhLiqCrys,SoftMPhys,CrossRMP93,HarrisonScience00} and for applications since they largely control the material properties. For the example of liquid crystals, most of the studies have focused on topological defects in the orientational ordered nematic phase of lyotropic or thermotropic liquid crystals and by now it has been well understood how to trigger them by external influences \cite{reznikov2000photoalignment,Guetal_PRL_2000,SaturnRingDefects,cladis1987dynamics,link1996simultaneous} and confinements \cite{lavrentovich1986phase,dammone2012,trukhina2008,VirNemConfGeom,Jackson2017,garlea2016finite}. In the layered smectic phase, defect characterization is more complex due to the additional positional ordering \cite{kleman2000grain,chen2009symmetry,liarte2015,Jeongetal_PRL_2012} but can be varied by confinement as well \cite{annulus,MonderkampPRL21}.

In recent years, active particles that are self-propelled intrinsically and relevant for biological swarms, motor proteins and biofilaments have been of tremendous interest \cite{Marchettietal_RMP_2013,Bechingeretal_RMP_2016,BerryRPP18,BowickPRX22}. These particles self-organize into fascinating ``active'' liquid crystalline phases which are qualitatively different from their passive counterparts in or near equilibrium and are governed by nonrelaxational dynamical processes that are the characteristics of far-from-equilibrium pattern-forming or chaotic systems \cite{CrossRMP93}. One important example of active matter systems involves active nematics showing orientational order. The dynamics of topological defects in active nematics have been analyzed to a large extent and a plethora of new phenomena were discovered (see, e.g., Refs.~\cite{SanchezNature12,GiomiPRL13,GiomiPTA14,Thampietal_EPL_2014,Thampietal_PTRSA_2014,Duclosetal_NP_2017,Sawetal_Nature_2017,Doostmohammadi_NC_2018,AnghelutaNJP2021,Keogh2021}). In particular, effects of confinement were explored for channels and capillaries \cite{OpathalagePNAS19,Hardouin_Comm_Physics_2019,Channel2,Channel3,Channel4,Channel5,Channel6,Channel7}, resulting in e.g., nontrivial dancing motion of defects \cite{Yeomans_Shendruk_SoftMatter_2017}. However, confinements different from channels were only rarely addressed \cite{curved1}. In parallel, effects of topology in the confinement have been studied for active nematics \cite{Keber_Science_2014,Ellis_NP_2018}. Also active layered smectic-like states which exhibit an additional positional order have been examined \cite{AdhyapakPRL13,Chenetal_PRL_2013}. These systems are modeled by aligning \cite{Romanczuketal_NJP_2016} and nonreciprocal \cite{Sahaetal_PRX_2020} interactions or nonlinear feedback \cite{Taramaetal_PRE_2019}. Due to the nontrivial coupling between orientational and positional degrees of freedom with the active driving force, active layered smectic systems and their defect dynamics are much more complex than their nematic counterparts. To the best of our knowledge, studies of defects in active smectics, including their dynamics and controllability by external constraints, are still sparse. Moreover, the effect of sharp cusps in the confinement shape has never been addressed for any active liquid crystalline or active pattern systems.

Here, we contribute to fill this gap and explore defect dynamics of active smectic systems by using an active phase field crystal (PFC) model \cite{Menzeletal_PRL_2013,menzel2014,Alaimoetal_NJP_2016,Alaimoetal_PRE_2018,Praetoriusetal_PRE_2018,Ophausetal_PRE_2018,Huangetal_PRL_2020,Ophausetal_PRE_2021} in a parameter setting where a traveling stripe phase is stable. We expose this state of self-propelled smectics to strong confinements where the smectic layer width is getting comparable to or not far from the confining length scales. A rich variety of boundary geometries with different topologies, including cavities and open or closed channels that are of various convex, concave, and cusped shapes, are considered. Our purpose here is to attain a systematic understanding of the complex dynamics of defects in the confined active smectics, particularly the effects of both geometry and topology. The confinements examined can be classified into two types of topology, closed cavity vs channel/ring (including closed rings and open channels with periodic boundary condition along open ends). Each of them includes different types of confining geometries with different number of sharp, singular cusps, either inward cusps (in epicycloid cavities or rings or cycloid open channel) or outward cusps (in hypocycloid cavities), or both combined (as in hypocycloid rings). Each type of topology also involves smooth boundaries with the lack of cusps (such as circular cavity, S-shaped open channel, and annulus).

In this study a wealth of nonequilibrium defected states are found in the evolution towards complex spatiotemporal patterns arising from a competition of activity and confinement. The mechanisms governing the dynamics of defects at large enough activity strongly differ from those of passive patterns and active nematics. The persistent defect dynamics identified here, especially the highly fluctuating defect state, goes far beyond the traditional classification familiar in passive systems, and is shown to be induced by the confining shape and topology of the cavity or channel and the degree of particle self-propulsion. This dynamical regime of high defect fluctuations occurs at intermediate activity as characterized by the perpetual process of defect creation-propagation-annihilation, showing as intermittently varying time stages with bursts of defects emerging in most of boundary confinements other than the smooth-boundary channels without cusp (i.e., S-shaped channel and annulus). In particular, the presence of cuspated boundaries induces or annihilates defects, which can be utilized to control the dynamics of defect creation and the quantitative behavior (oscillatory vs damped) of time correlation of defect density. Our predictions can be verified for confined dense vibrated granular rods \cite{narayan2006nonequilibrium,DeseignePRL10,armas2020} or self-propelled colloidal Janus particles \cite{VutukuriSoftMatt16,WykesSoftMatt16} exposed to strong confinements \cite{annulus}.

\section*{Results and Discussion}

\subsection*{Model}
We describe the evolution of active smectics under confinement based on a 
continuum density-field theory, i.e., the active PFC model which can be derived from dynamical
density functional theory \cite{Menzeletal_PRL_2013,menzel2014} and also from a particle-based
microscopic description \cite{Huangetal_PRL_2020}. It reads
\begin{eqnarray}
  && \frac{\partial \psi}{\partial t} = \nabla^2 \frac{\delta \mathcal{F}}{\delta \psi}
  - v_0 \bm{\nabla} \cdot \mathbf{P}, \label{eq:psi}\\
  && \frac{\partial \mathbf{P}}{\partial t} = \left ( \nabla^2 - D_r \right )
  \frac{\delta \mathcal{F}}{\delta \mathbf{P}} - v_0 \bm{\nabla} \psi, \label{eq:P}
\end{eqnarray}
where $\psi$ is the particle density variation field, the polarization $\mathbf{P}$ 
represents the local orientation vector field, $v_0$ measures the strength of particle 
self-propulsion, and $D_r$ is the rotational diffusion constant. The above dynamical 
equations have been rescaled, with a diffusive timescale and a length scale set via 
the pattern periodicity. The model considers conserved dynamics for 
$\psi$, as seen in Eq.~(\ref{eq:psi}),
and thus the average density $\psi_0$ remains unchanged during the system evolution.
We set $\mathcal{F} = \mathcal{F}_{\rm aPFC} + \mathcal{F}_{\rm anch}$,
where $\mathcal{F}_{\rm aPFC}$ is the rescaled free energy functional of active PFC
\cite{Menzeletal_PRL_2013,menzel2014}
\begin{equation}
  \mathcal{F}_{\rm aPFC} = \int d\mathbf{r} \left \{ \frac{1}{2} \psi \left [ \epsilon
    + \left ( \nabla^2 + q_0^2 \right )^2 \right ] \psi %- \frac{g}{3} \psi^3
    + \frac{1}{4} \psi^4  + \frac{C_1}{2} \vert\mathbf{P}\vert^2 \right \},
  %+ \frac{C_4}{4} \vert\mathbf{P}\vert^4 \right \},
\end{equation}
with $\epsilon<0$, the characteristic wave number
$q_0=1$ after rescaling, and $C_1>0$ tending to suppress any spontaneous ordering 
of orientational alignment. $\epsilon$ and the average density $\psi_0$ are chosen to give rise to the resting or traveling active smectic phase \cite{Menzeletal_PRL_2013}. 

We represent the effect of boundary confinement via an anchoring energy
\begin{equation}
  \mathcal{F}_{\rm anch} \!=\! \int \! d\mathbf{r} \frac{V_b(\mathbf{r})}{2} \left [
    (\psi-\psi_b)^2 + \vert \mathbf{P}-\mathbf{P}_b \vert^2
    + \left ( \hat{\mathbf n} \cdot \bm{\nabla} \psi \right )^2 \right ], \label{eq:F_anch}
\end{equation}
to effectively satisfy both Dirichlet and Neumann boundary conditions $\psi=\psi_b$,
$\mathbf{P}=\mathbf{P}_b$, and $\hat{\mathbf n} \cdot \bm{\nabla} \psi = 0$ (with $\hat{\mathbf n}$
the local unit normal) at any implicitly defined domain boundary, with 
\begin{equation}
  V_b(\mathbf{r}) = \frac{V_{b0}}{2}
  \left [ 1 + \tanh \left ( \frac{r_s(\mathbf{r})}{\Delta} \right ) \right ],
  \label{eq:Vb}
\end{equation}
where $V_{b0}$ gives the anchoring strength, $r_s(\mathbf{r})$ is the signed distance function to the domain boundary (with $r_s<0$ inside the domain and $>0$ outside), and $\Delta$ sets the thickness of boundary interface. This approach we develop here combines an approximation of domain interface energy for imposing the boundary conditions and a setup of boundary via $V_b(\mathbf{r}) = V_{b0} \phi(\mathbf{r})$ with an auxiliary phase field function $\phi(\mathbf{r})$ used in the diffuse domain method \cite{LiCMS09} to control the confinement geometry implicitly. More details, including the specific analytical forms for different geometries of the cavities or channels simulated, are given in the Methods section. Among them, the geometry of epicycloid or hypercycloid with integer $n$ cusps is described as a closed plane curve formed by the rolling of a small circle of radius $b$ on the outside or inside of another larger fixed circle of radius $a=nb$, respectively, with their parametric equations given in Eqs.~(\ref{eq:epicycloid}) and (\ref{eq:hypocycloid}) of the Methods section. Equation (\ref{eq:F_anch}) produces the condition of planar anchoring as found in experiments. Its last term is analogous to the Rapini-Papoular form of surface potential \cite{SoftMPhys,VirNemConfGeom}. In our simulations (starting from random initial conditions), we set $(\epsilon, \psi_0, D_r, C_1) = (-0.98, 0, 0.5, 0.2)$ in the strong segregation regime of stripe phase, and $(V_{b0}, \Delta, \psi_b, \mathbf{P}_b) = (1, 0.1, 0, 0)$ for the cavity or channel boundary setup. We have tested stronger anchoring strength with larger value of $V_{b0}$ and found that larger self-propulsion strength $v_0$ would then be needed to overcome the stronger boundary confinement particularly for defect nucleation, while the corresponding results obtained are qualitatively similar.

\subsection*{Defects dynamics in closed cavities}

\begin{figure}
    \centering
    \includegraphics[width=0.75\textwidth]{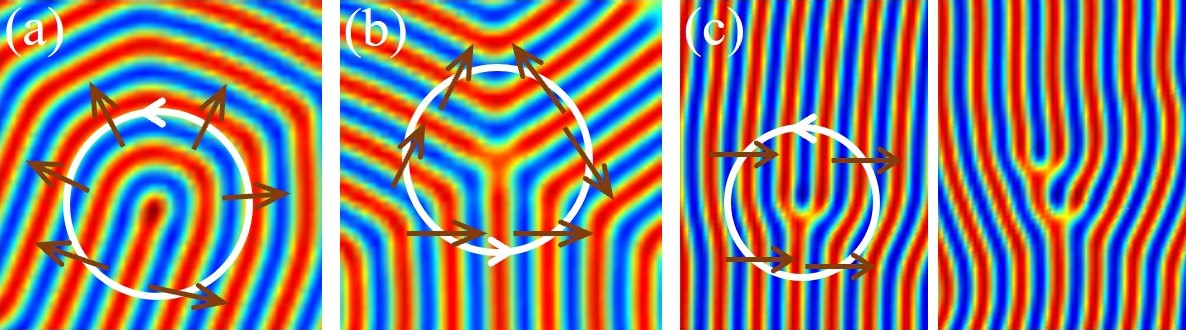}
    \caption{Topological defects in active smectics. Typical topological defects found in simulations of active smectic pattern as determined by the density field profile, including (a) $+1/2$ and (b) $-1/2$ disclinations, and (c) a single and a doublet of edge dislocations. Some local normal directions of smectic layers are indicated by arrows, and an integral of their orientational angle $\theta_s$ over a counterclockwise closed-path loop around the defect core (white-circled) determines its topological charge via $\oint d\theta_s / 2\pi = \pm 1/2, 0$ for disclinations and dislocations respectively. (Note that angles $\theta_s$ with a $\pi$ difference are equivalent to each other.)}
    \label{fig:defects_schematic}
\end{figure}

\begin{figure*}
  \centerline{\includegraphics[width=\textwidth]{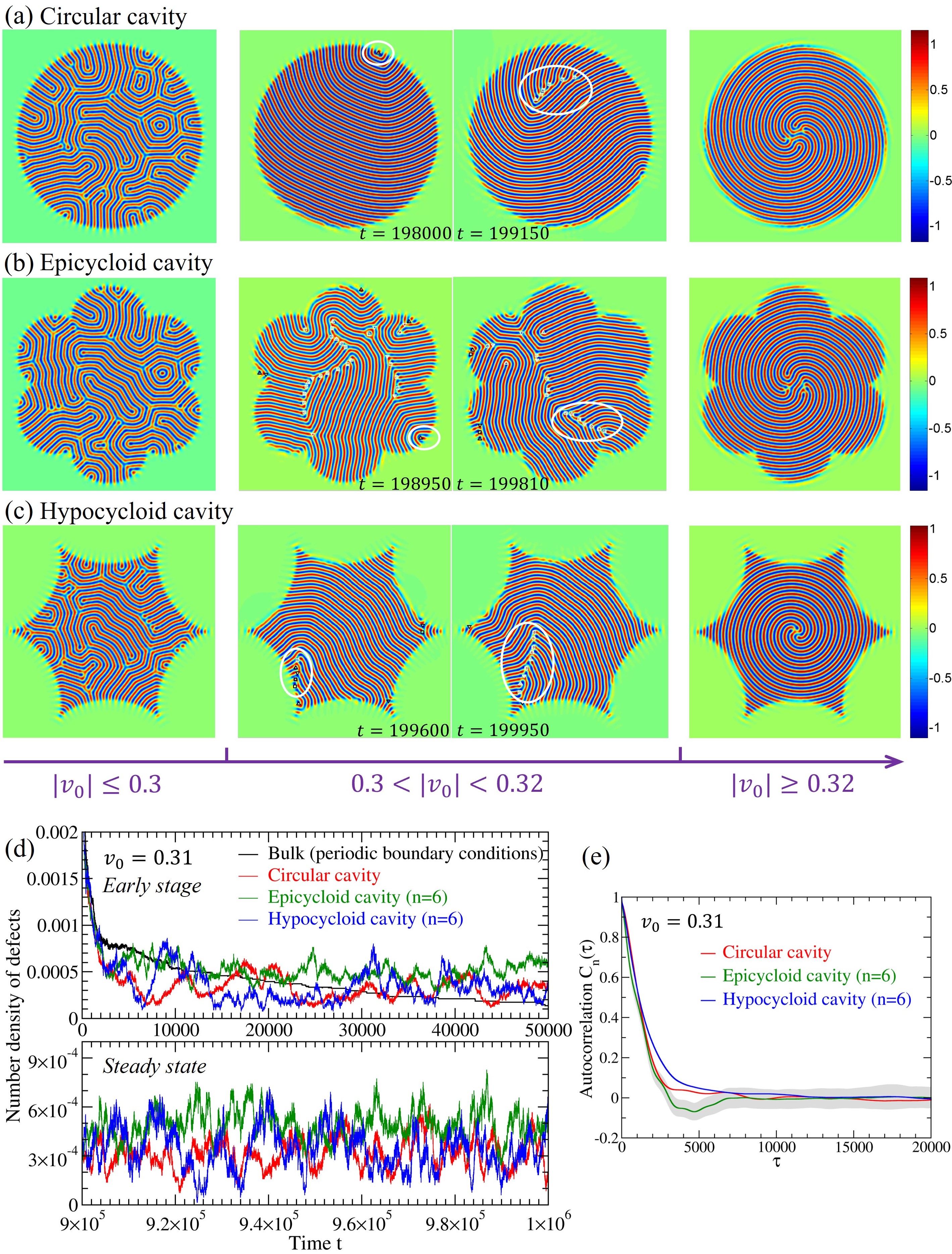}}
  \caption{Three dynamical regimes of active smectics defect evolution in closed cavities.
    (a)--(c) Transitions between pinned, highly fluctuating (annihilation-nucleation),
    and self-rotating defect states with the increase of self-propulsion strength $v_0$, in 
    (a) circular, (b) epicycloid, and (c) hypocycloid cavities. 
    Each regime is represented by sample snapshots of the spatial 
    profiles of density field $\psi$ obtained from simulations,
    with the $\psi$ scale labelled by the color bars.
    %with system size of $256 \times 256$ grid points. 
    In the mid panels the circled regions highlight the time-evolving process of 
    boundary-induced defect generation. The bulk defects inside are labeled by 
    white symbols, and boundary defects by black ones. Among them the square
    symbols indicate dislocations and the up or down triangles indicate $\pm 1/2$
    disclinations respectively. 
    (d) Sample time variation of defect number density in the fluctuation regime at 
    $v_0=0.31$ (for system size of $512 \times 512$ grid points). 
    (e) The corresponding normalized time correlation $C_n(\tau)$
    of defect density, calculated over $t=10^5-10^6$ and averaged over 80
    simulations for each cavity. For a better illustration only the error-bar band
    for epicycloid of $n=6$ cusps is shown, while those for other cases are of similar range.}
  \label{fig:transitions}
\end{figure*}

The emergence of topological defects, including disclinations, dislocations, and 
grain boundaries, is observed in our simulation systems 
(see Fig.~\ref{fig:defects_schematic} for some typical
smectic defects, similar to those found in passive smectic or stripe-pattern systems 
\cite{chaikin_lubensky_1995all,PhLiqCrys,SoftMPhys,CrossRMP93,HarrisonScience00}). 
Their complex behaviors of dynamical evolution is further complicated by the effects 
of active self-driving and boundary confinement. In the confined cavities of different 
geometries, our results presented in Fig.~\ref{fig:transitions} show that the evolution 
of these defects in active smectics is governed by three intrinsically different 
dynamical regimes. At weak enough self-propulsion strength $v_0$ 
(first column of Figs.~\ref{fig:transitions}(a)--\ref{fig:transitions}(c)), the
stripes (smectic layers) remain perpendicular to the boundary, showing planar 
anchoring with tangential alignment of constituent particle orientations. The defects 
emerging from the early stage of system evolution become mostly pinned, with extremely 
slow local dynamics, similar to the glassy state observed in strongly segregated 
passive stripe patterns showing no long range orientational order as a result of 
defect pinning by the pattern-periodicity induced potential barrier \cite{BoyerPRE02}.

In the other limit of large enough $v_0$, the particle self-propulsion completely overcomes
the defect pinning barrier, facilitating the fast annihilation of defects as accelerated
by the effect of self-driving on the smectic layer alignment.
As shown in the last column of Figs.~\ref{fig:transitions}(a)--\ref{fig:transitions}(c),
large particle activity also enables the overcoming of boundary anchoring constraint,
leading to the violation of local planar anchoring, and depending on the boundary geometry,
even to (partially) homeotropic anchoring. At late stage such a strong self-driving induces the
rotation of smectics inside the cavity, either clockwise or counterclockwise, and hence the
persistent self-rotation of the remaining multi-core spiral defects trapped at the cavity center.

The transition between these different dynamical regimes occurs within a narrow range of the self-propulsion strength $v_0$ (middle column of Figs.~\ref{fig:transitions}(a)--\ref{fig:transitions}(c)). To understand this transition between localized and traveling smectic patterns, we focus on the bulk state and neglect the boundary anchoring energy. This allows us to rewrite Eqs.~(\ref{eq:psi}) and (\ref{eq:P}) via defining a local
polarization divergence field $S=\bm{\nabla} \cdot \mathbf{P}$, i.e.,
\begin{eqnarray}
  && \frac{\partial \psi}{\partial t} = \nabla^2 \left [ \epsilon \psi + \left ( \nabla^2 + q_0^2 \right )^2 \psi
  + \psi^3 \right ] - v_0 S, \label{eq:psi_S}\\
  && \frac{\partial S}{\partial t} = C_1 \left ( \nabla^2 - D_r \right ) S
  - v_0 \nabla^2 \psi. \label{eq:S}
\end{eqnarray}
Working in a comoving frame with $\mathbf{r} \rightarrow \mathbf{r}' = \mathbf{r} - \mathbf{v}_m t$
and assuming $\psi = \psi(\mathbf{r} - \mathbf{v}_m t)$ and $S = S(\mathbf{r} - \mathbf{v}_m t)$
in the nonequilibrium steady state of pattern dynamics, where $\mathbf{v}_m$ is the 
migration velocity of smectic layers, we can obtain the equations governing the Fourier
components $\hat{\psi}_{\mathbf q}$ and $\hat{S}_{\mathbf q}$ of the $\psi$ and $S$ fields as
\begin{eqnarray}
  && i(\mathbf{q} \cdot \mathbf{v}_m) \hat{\psi}_{\mathbf q} = q^2 \left \{
  \left [ \epsilon + \left ( q^2 - q_0^2 \right )^2 \right ] \hat{\psi}_{\mathbf q}
  + \psi^3 \vert_{\mathbf q} \right \} + v_0 \hat{S}_{\mathbf q}, \label{eq:psiq}\\
  && i(\mathbf{q} \cdot \mathbf{v}_m) \hat{S}_{\mathbf q} = C_1 \left ( q^2 + D_r \right )
  \hat{S}_{\mathbf q} - v_0 q^2 \hat{\psi}_{\mathbf q}. \label{eq:Sq}
\end{eqnarray}
In one-mode approximation the density field for a perfect stripe phase is given by
$\psi = A \exp[i \mathbf{q}_s \cdot (\mathbf{r} - \mathbf{v}_m t)] + \textrm{c.c.}
= A \exp(i \mathbf{q}_s \cdot \mathbf{r}') + \textrm{c.c.}$, where $\mathbf{q}_s$ is 
the selected wave vector of the pattern and $A=A_0 e^{i\phi_0}$ with a constant phase
$\phi_0$. Here we have assumed that the model parameters (including $\epsilon$, $v_0$, $C_1$, and $D_r$) are within the range of the traveling state of active smectic pattern examined in this work.
For mode $\mathbf{q}_s$ Eqs.~(\ref{eq:psiq}) and (\ref{eq:Sq}) then become
\begin{eqnarray}
  && \left [ \epsilon + \left ( q_s^2 - q_0^2 \right )^2 + \frac{v_0^2} 
  {C_1 \left ( q_s^2 + D_r \right ) - i\mathbf{q}_s \cdot \mathbf{v}_m}
  - \frac{i\mathbf{q}_s \cdot \mathbf{v}_m}{q_s^2} \right ] A_0 + 3A_0^3 =0,~~ \label{eq:A0}\\
  && \hat{S}_{\mathbf{q}_s} = \frac{v_0 q_s^2 A}{ C_1 \left ( q_s^2 + D_r \right ) 
  - i\mathbf{q}_s \cdot \mathbf{v}_m}. \label{eq:S_qs}
\end{eqnarray}
Separating real and imaginary parts of Eq.~(\ref{eq:A0}) leads to the following solutions
for velocity $\mathbf{v}_m$ and amplitude $A_0$: When $\vert v_0 \vert \leq v_{0c}$, we have
$\mathbf{v}_m=0$ corresponding to localized stripe patterns, with
\begin{equation}
    A_0^2(\mathbf{v}_m=0) = -\frac{1}{3} \left [ \epsilon + \left ( q_s^2 - q_0^2 \right )^2 
    + \frac{v_0^2}{C_1 \left ( q_s^2 + D_r \right )} \right ],
\end{equation}
while at large enough activity $\vert v_0 \vert > v_{0c}$, the pattern travels with nonzero 
velocity $\mathbf{v}_m$ as determined by
\begin{eqnarray}
    && \mathbf{q}_s \cdot \mathbf{v}_m = q_s \left ( v_0^2 - v_{0c}^2 \right )^{1/2}, \\
    && A_0^2 = -\frac{1}{3} \left [ \epsilon + \left ( q_s^2 - q_0^2 \right )^2 
    + C_1 \left ( 1 + D_r/q_s^2 \right ) \right ],
\end{eqnarray}
where $v_{0c}$ is the critical threshold of activity for the transition, as given by
\begin{equation}
    v_{0c} = C_1 (q_s^2 + D_r)/q_s. \label{eq:v0c}
\end{equation}
For a nonpotential, nonrelaxational system like the active system studied here, the selected wave number $q_s$ of an ordered pattern in the long-time steady state cannot be determined from free energy minimization and is difficult to identify through analytical calculations \cite{CrossRMP93}. In the active PFC model the value of $q_s$ is expected to be near $q_0$ as also found in our numerical simulations.
Substituting the parameter values used in our simulations (i.e., $C_1=0.2$ and $D_r=0.5$)
and approximating $q_s \sim q_0=1$, we get $v_{0c} = 0.3$
from the above analytic result, the same as what has been found in simulations and 
presented in Figs.~\ref{fig:transitions}(a)--\ref{fig:transitions}(c). 

The direction of pattern traveling velocity $\mathbf{v}_m$ and the orientation of the selected 
wave vector $\mathbf{q}_s$ (which tend to locally align with each other) highly depend 
on the initial and boundary conditions, and could vary in different parts of the 
system due to the effect of boundary confinement, as seen in our numerical simulations 
(e.g., Fig.~\ref{fig:transitions} and Supplementary Movies 1-7).

Defect dynamics in these different regimes (localized and traveling smectic 
patterns and their transition) reveal the competition between rigid boundary
confinement restricting the local smectic orientation and the tendency of bulk alignment 
of stripes \cite{CrossPRA82,GreensidePRA84}. An interesting type of dynamics occurs 
in the transition regime when such two incompatible boundary and bulk effects 
are of comparable strength, giving rise to a highly fluctuating state (middle column of 
Figs.~\ref{fig:transitions}(a)--\ref{fig:transitions}(c)). Although the planar boundary 
anchoring is maintained at very early stage, the deviation occurs
at later times as caused by self-propulsion, leading to local distortion of stripes inside 
the cavity as a result of confinement-alignment competition. Importantly, in addition to the
annihilation of defects (including dislocations and disclinations, majorities of which 
occur at cavity boundaries), new defects can nucleate from the boundary, propagating into 
the bulk, evolving and generating a subsequence of more new defects like a chain effect, 
as seen in the circled regions of Fig.~\ref{fig:transitions} and 
Supplementary Movies 1-7 for epicycloid and hypocycloid cavities. 
This results in the repeated succession of tranquil and active time stages in terms of 
defect density and dynamics, with some examples for circular and 6-cusp epicycloid and 
hypocycloid cavities given in Fig.~\ref{fig:transitions}(d). In contrast to the fully 
bulk state without any boundary confinement (thus with the absence of defect generation) 
which shows a monotonic time decay of defect number, the cavity confinement induces an
intermittency-type behavior with seemingly irregular bursts of number of defects. The 
boundary cusps appear to enhance the creation of new defects, yielding higher defect 
density peaks, as compared to the smooth boundary of circular cavity. 
It is noted that the overall system activity is governed by $v_0$
which is kept unchanged for different confinement geometries and cusp number, although
there would be local variations of effective activity as a result of complex nonlinear 
defect dynamics particularly defect creation and annihilation.

The property of this transition zone with dynamical fluctuations of defects can be further
quantified through the normalized time autocorrelation function of the defect number density
$n_d$, i.e.,
\begin{equation}
  C_n(\tau) = \frac{\langle \left ( n_d(t+\tau) - \langle n_d \rangle \right )
    \left ( n_d(t) - \langle n_d \rangle \right ) \rangle}
  {\langle \left ( n_d(t) - \langle n_d \rangle \right )^2 \rangle},
\end{equation}
where the averages are conducted over a long time series in the steady state (e.g., $t=10^5-10^6$
in our calculations) for each simulation run, assuming ergodicity of the corresponding probability
measure \cite{CrossRMP93}. Some results of $C_n(\tau)$ %(each averaged over 80 independent runs)
are presented in Fig.~\ref{fig:transitions}(e), showing a decay behavior for circular and 6-cusp
hypocycloid cavities. Interestingly, for epicycloid cavity with $n=6$ cusps a weak oscillation
around a negative minimum correlation (near time scale $\tau_m \sim 5000$) appears, implying
a correlated behavior between the burst (active) and low-number (tranquil) regimes of defect
density and dynamics.

\begin{figure}
  \centerline{\includegraphics[width=0.85\textwidth]{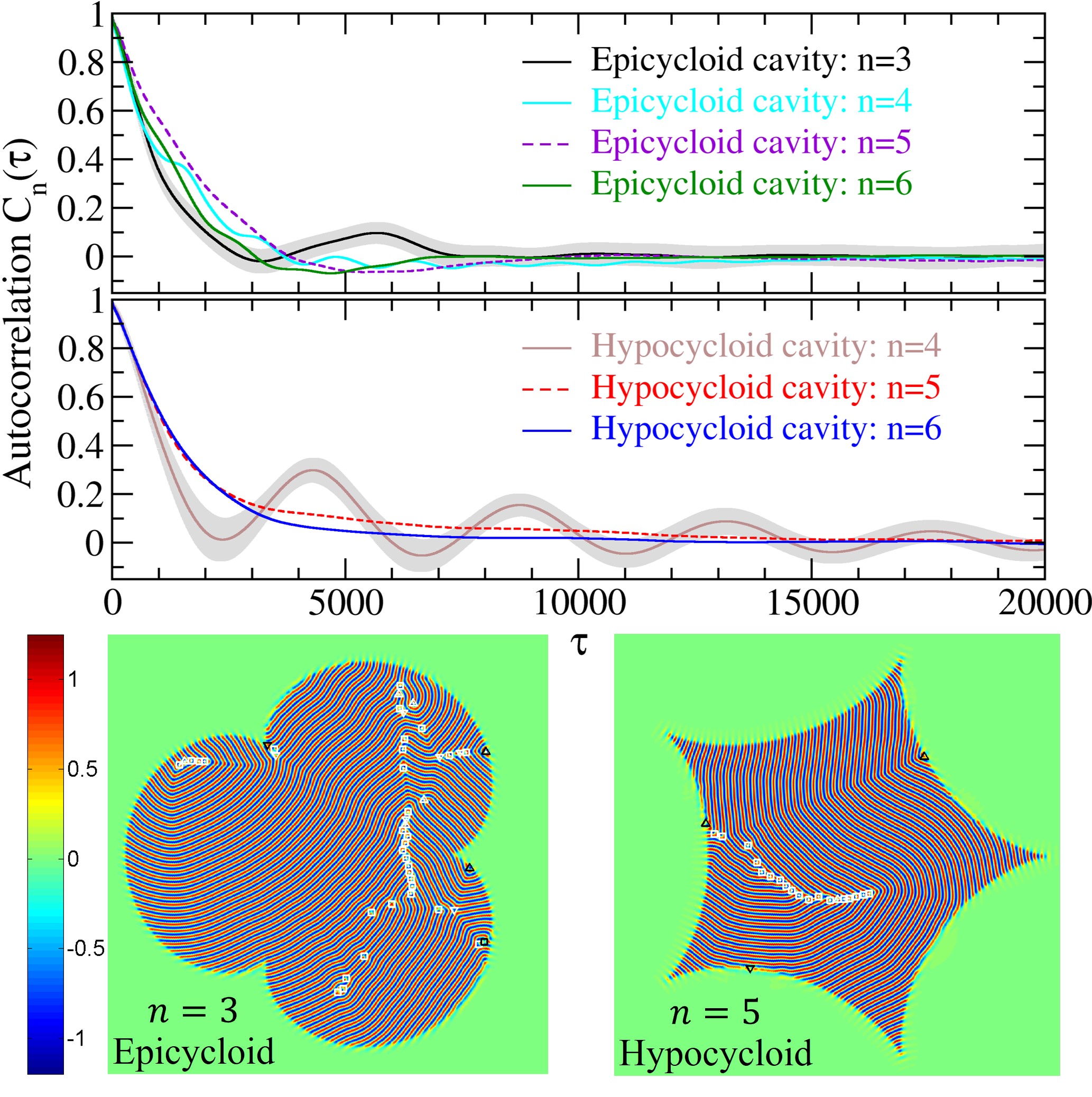}}
  \caption{Time correlation of defect density in closed cavities.
    Autocorrelation function $C_n(\tau)$ of defect number density for various 
    epicycloid and hypocycloid closed cavities at $v_0=0.31$. Also shown are 
    two sample snapshots of the $\psi$ field spatial profile, 
    with the scale of $\psi$ indicated by the color bar
    and the same symbol labeling of defects as that in Fig.~\ref{fig:transitions}.}
  \label{fig:cavities_Corr}
\end{figure}

For a given geometric type of confinement, the behavior of defect autocorrelation can be
qualitatively changed through different number of boundary cusps. As shown in
Fig.~\ref{fig:cavities_Corr}, for epicycloid cavities decreasing the cusp number $n$ from
$n=6$ to $3$ leads to the variation of $C_n(\tau)$ 
from local negative minimum to positive maximum of
correlation. A more dramatic change occurs when lowering the cusp number of hypocycloid
cavities. When $n=5$ and $6$ a damped time correlation of defect density is observed, while
the $n=4$ (i.e., astroid) cavity is featured by an oscillatory behavior (within the statistical
error) of time correlation, indicating a cyclic state of defect density variation with periodic
creation and annihilation of defects over a characteristic time period $\tau_T \sim 4300$.
In this cyclic defect dynamics, although generally the spatial locations of boundary defect
nucleation and annihilation seem uncorrelated, statistically the periodicity in autocorrelation
$C_n(\tau)$ can be attributed to the propagation of defects between different sides of boundary
within the defect creation-annihilation time interval.

\begin{figure}
    \centering
    \includegraphics[width=0.85\textwidth]{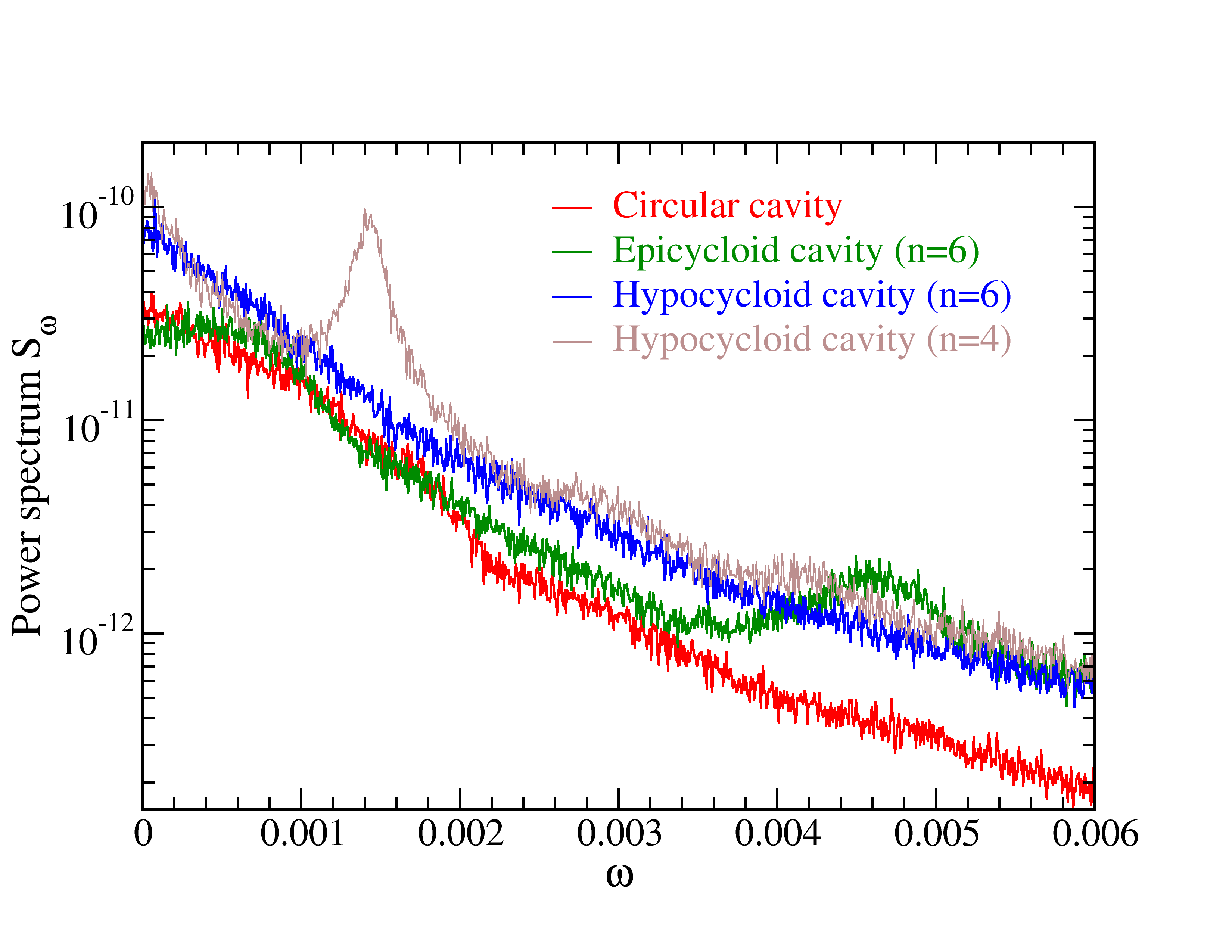}
    \caption{Power spectrum for time variation of defect density in closed cavities. Power spectrum $S_\omega$ of the defect number density for circular, 6-cusp epicycloid, and 4- and 6-cusp hypocycloid cavities at $v_0=0.31$, each averaged over 80 simulations for $t=10^5-10^6$.}
    \label{fig:cavities_Somega}
\end{figure}

Figure \ref{fig:cavities_Somega} gives the corresponding power spectrum of the defect number density for some closed cavities with different number of cusps. It has been averaged over independent simulation runs as calculated via
\begin{equation}
    S_\omega = \frac{1}{L_t} \langle \vert\hat{n}_d(\omega)\vert^2 \rangle,
\end{equation}
where $\hat{n}_d(\omega)$ is the temporal Fourier transform of defect density $n_d(t)$ and $L_t$ is the length of $n_d$ time series which needs to be long enough particularly for non-periodic variation of $n_d$ (in our calculations a time range of $t=10^5-10^6$ for each run is used, with 80 different runs). The inverse Fourier transform of this power spectrum $S_\omega$ gives the unnormalized time autocorrelation function $\langle n_d(t+\tau) n_d(t) \rangle$, according to the Wiener-Khinchin theorem. An oscillatory behavior of autocorrelation $C_n(\tau)$ then corresponds to a single high peak in the power spectrum plot, as shown in Fig.~\ref{fig:cavities_Somega} for the $n=4$ hypocycloid cavity with a peak located around $\omega \sim 2\pi/\tau_T = 0.00146$, consistent with its $C_n(\tau)$ result given in Fig.~\ref{fig:cavities_Corr}.

The complex behaviors of defect dynamics revealed above for active smectic systems are
governed by intrinsically different mechanisms compared to those examined in previous
works of defect dynamics in stripe or convection-roll patterns of passive systems 
\cite{CrossRMP93}. Most of those passive model systems, either potential or nonpotential, 
are governed by nonconserved dynamics, such as the original or generalized Swift-Hohenberg 
equations with or without the coupling to hydrodynamic mean flow and vertical vorticity
\cite{SiggiaPRA81,PomeauPRA83,TesauroPRA86,Pismen99,BoyerPRE02,VitralPRF20}. 
For conserved dynamics as studied here, the treatments are more complicated and some 
related developments for the study of dislocation motion in passive crystalline systems 
were available only recently \cite{SkaugenPRB18,SalvalaglioPRL21}. The considered approach was based on amplitude equation 
expansion in the weakly nonlinear regime and applied to the pattern 
near onset. This is different from the active system studied here which is
nonpotential and nonrelaxational and is far from onset with strong segregation. The 
corresponding amplitude equation formulation and the related analysis for strongly segregated patterns would be much more involved and need the incorporation of nonadiabatic effects for lattice pinning \cite{BoyerPRE02,HuangPRE13,HuangPRE16}. 

The defect dynamics here also differ from those known in active nematics, where they strongly depend on the type of the defects and the interactions between them \cite{SanchezNature12,GiomiPRL13,GiomiPTA14,Doostmohammadi_NC_2018,BowickPRX22}. In active smectics the defect type and the interactions between individual defects are only of secondary effect for the parameter range and the corresponding dynamical regimes examined in this work. At low activity (i.e., $v_0 \leq v_{0c}$ with the threshold $v_{0c}$ determined by Eq.~(\ref{eq:v0c})), since the system examined here is in the strong segregation regime (with $\epsilon=-0.98$), the defects are pinned to the underlying periodic structure of the stripe pattern (similar to that of a passive potential system exhibiting a stripe phase in Ref.~\cite{BoyerPRE02}); thus the driving forces caused by the intrinsic interactions between defects (between either different dislocations or disclinations) are much smaller than the pinning force and do not overcome the pinning barrier. When $v_0 > v_{0c}$ at large enough activity which is the main focus here, the active self-driving force overcomes the pattern pinning effect and when combined with the effect of boundary confinement, dominates the evolution and motion of all the defects as it clearly exceeds the pinning force and hence far exceeds the defect-defect interactions which now play a secondary role. This can be observed in numerical simulations (see Supplementary Movies 1-7), where at the leading order the motion of defects mostly follows the traveling of local stripe layers as driven by the self-propulsion, accompanied by further defect generation/splitting or annihilation through the coupling to rigid boundary confinement and the effect of large activity.

\begin{figure}
  \centerline{\includegraphics[width=\textwidth]{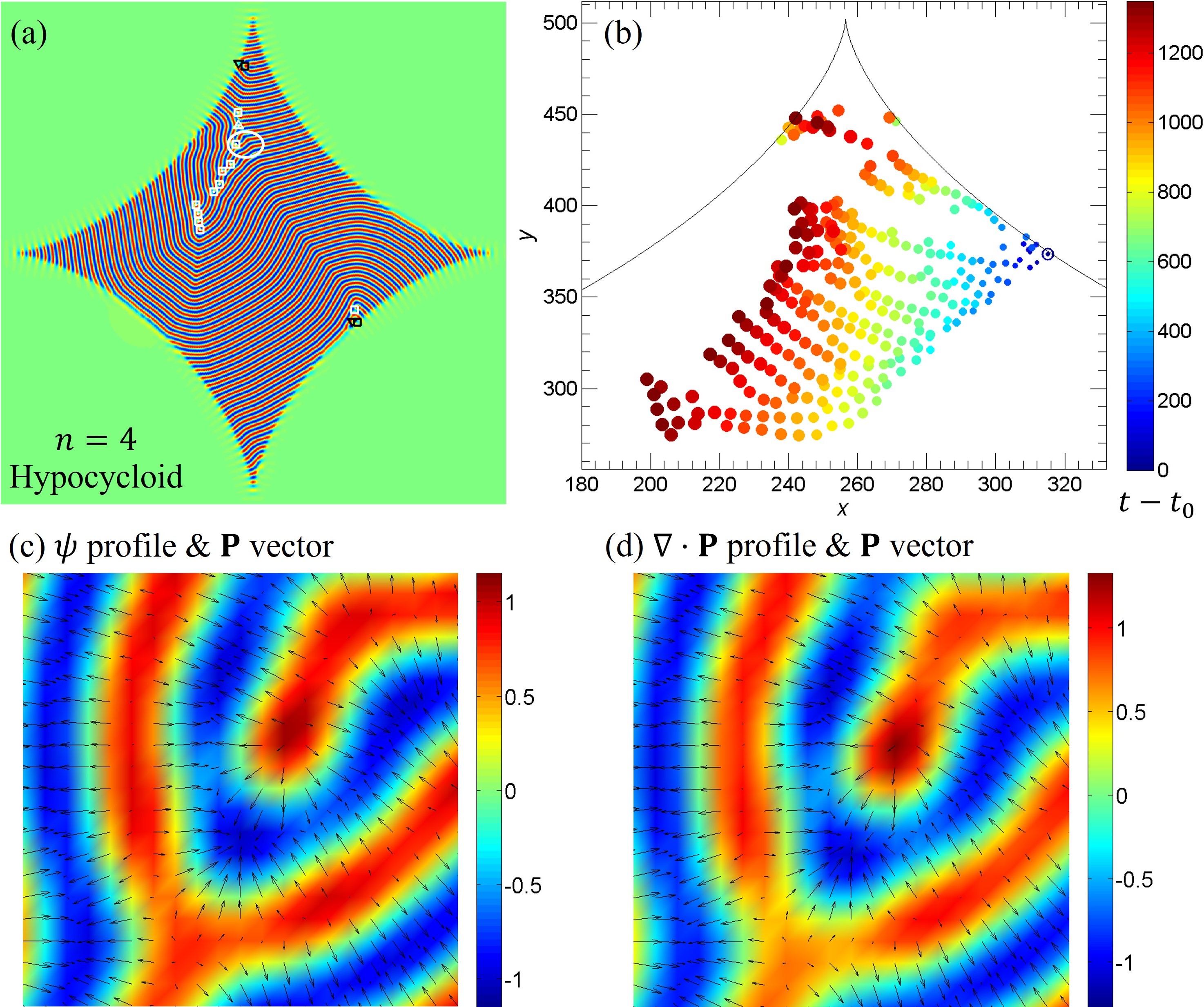}}
  \caption{Spatial profiles of $\psi$ and $S$ fields and defect trajectories in hypocycloid cavity.
  (a) Sample simulation snapshot of the density field $\psi$ 
  profile for a 4-cusp hypocycloid (astroid) cavity at $v_0=0.31$, with the labeling of 
  defects. (b) Trajectories of defects in the upper part of the cavity, starting from a 
  single defect (circled on the right) nucleated at the boundary at time $t_0=198650$
  up to the snapshot of (a) at $t=2 \times 10^5$. Larger symbol size corresponds
  to later time, and samples for different time are also color-coded according to 
  the color bar given on right. The hypocycloid boundary curve is also plotted. The
  corresponding time evolution of the pattern and defects can be found in Supplementary
  Movie 5. (c) Enlarged portion of the circled region in (a), with black arrows
  indicating the polarization vector field $\mathbf{P}$ and the 
  color bar showing the scale of density field $\psi$. (d) The corresponding spatial
  profile of the divergence field $S=\bm{\nabla}\cdot\mathbf{P}$, with $\mathbf{P}$
  vector also shown as arrows and the scale of $S$ field indicated
  by the color bar. Note a small phase shift of this $S$ field profile
  with respect to the $\psi$ profile in (c), which corresponds to a nonzero migration
  velocity.}
  \label{fig:trajectory_divP}
\end{figure}

Some sample trajectories of defect motion are illustrated in Fig.~\ref{fig:trajectory_divP} for an $n=4$ hypocycloid (astroid) cavity at $v_0=0.31$. The dynamics starts from a single defect created at the upper-right boundary of the cavity; then a chain of new defects is generated sequentially during the traveling into the bulk (see Fig.~\ref{fig:trajectory_divP}(b) and the corresponding Supplementary Movie 5). This procedure of new defect generation is caused by the active driving and local distortions of the smectic layers which depend on the specific geometry and topology of the boundary confinement (with more related studies given in the next section), and their motion is subjected to the flowing of local stripes as observed in Supplementary Movie 5.

The self-driving force for the time evolution of density field and the resulting 
defect dynamics is determined by the local polarization divergence term 
$v_0 \bm{\nabla}\cdot\mathbf{P}$ as can be seen from Eq.~(\ref{eq:psi}) or 
(\ref{eq:psi_S}). As an example, in Fig.~\ref{fig:trajectory_divP} we show the spatial 
distribution of the polarization field $\mathbf{P}$ and its divergence field 
$S=\bm{\nabla}\cdot\mathbf{P}$ (which is in turn coupled to the local variation 
of density field through $v_0 \nabla^2 \psi$ as given in Eq.~(\ref{eq:S})) 
around a defect core in a four-cusp astroid cavity. As illustrated in 
Fig.~\ref{fig:trajectory_divP}(c), the vector field $\mathbf{P}$ represents the 
distribution of local orientational order of active-driving directions, with the net 
self-propulsion determined by the asymmetric distribution of the $\mathbf{P}$ field 
surrounding the local density peak of $\psi$ and the corresponding net orientation 
of $\mathbf{P}$ \cite{Menzeletal_PRL_2013,menzel2014}. The information of self-driving
can then be obtained from the local spatial gradients of $\mathbf{P}$, with 
$\bm{\nabla}\cdot\mathbf{P}$ being of similar pattern as the density field
$\psi$ (see Fig.~\ref{fig:trajectory_divP}(d)). The only difference is a nonzero 
phase shift between them giving the effect of self-propulsion. This can be also 
seen from Eq.~(\ref{eq:Sq}) or (\ref{eq:S_qs}) showing their Fourier components 
($\hat{\psi}_{\mathbf q}$ and $\hat{S}_{\mathbf q}$) being proportional to 
each other but with a phase difference when the net migration velocity 
$\mathbf{v}_m \neq 0$.

\subsection*{Topology-dependent defect dynamics
and cusp-induced mechanism of defect generation}

\begin{figure}
  \centerline{\includegraphics[width=0.9\textwidth]{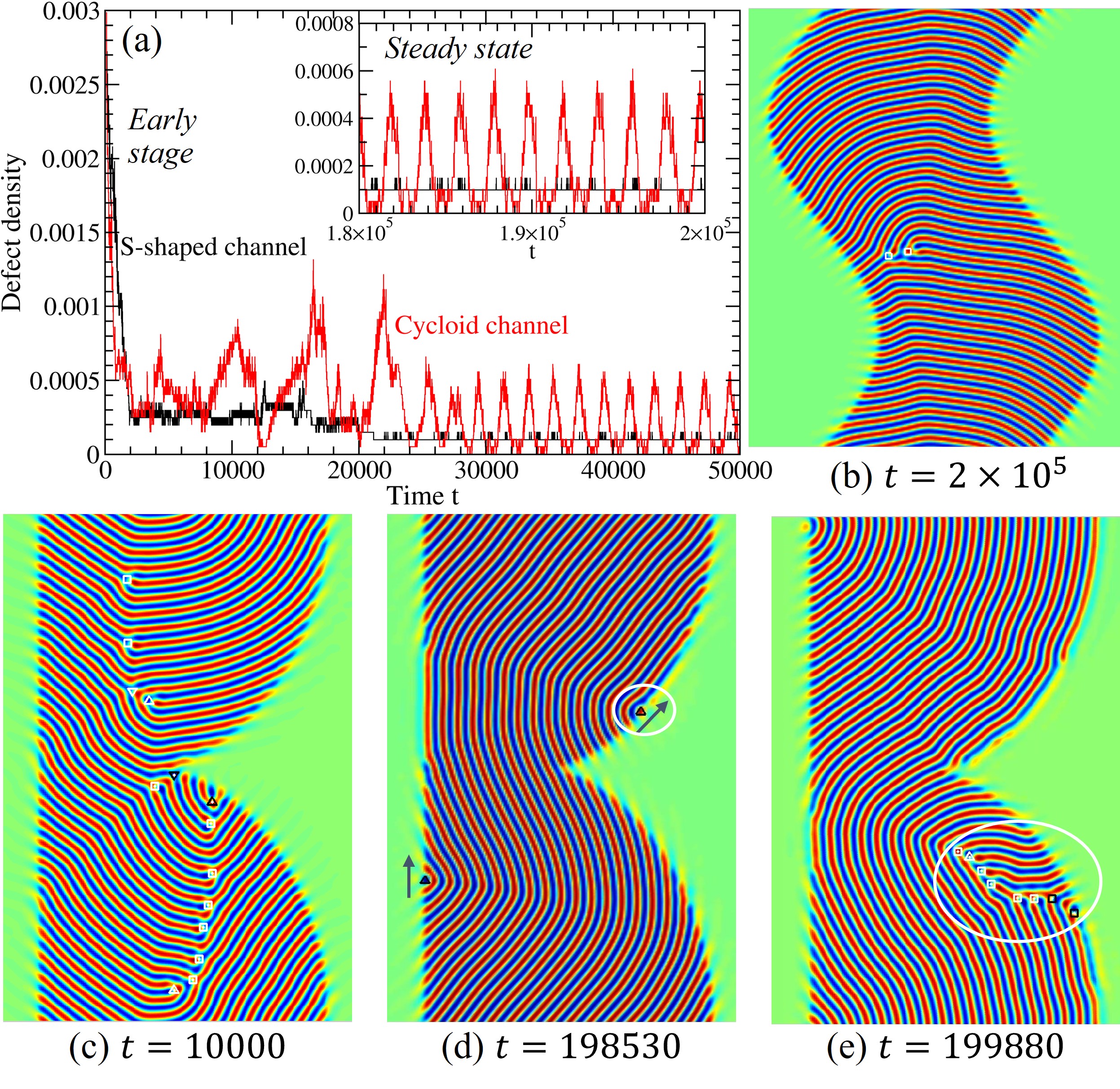}}
  \caption{Time evolution of defects in open channels. (a) Time evolution of defect number density in two types of open channels at 
    $v_0=0.31$ (with periodic boundary condition along the vertical open 
    ends). (b)--(e) Snapshots of $\psi$ profiles at different times 
    for (b) S-shaped and (c)--(e) cycloid channels.}
  \label{fig:channels_open}
\end{figure}

To further understand the mechanism of defect creation and correlation and hence the
accessibility for varying defect dynamics, we examine the process of defect flow 
for a different topology including two types
of open channels, the S-shaped channel with smooth boundary and the cycloid channel with a
single cusp (see Fig.~\ref{fig:channels_open}, noting the periodic boundary condition along
the vertical open ends). No new defects are nucleated from the smooth boundary of S-shaped
channel and the defect number decreases with time, with few defects left at late stage, as seen in
Figs.~\ref{fig:channels_open}(a) and \ref{fig:channels_open}(b). In contrast, during the flow
of stripes in the cycloid channel (Figs.~\ref{fig:channels_open}(c)--\ref{fig:channels_open}(e)),
the cusp singularity enhances the local distortion of the smectic layers, thus enabling the
formation of defects at boundary (not necessarily at the cusp location). Further distortion
during flow can facilitate the defect motion into the bulk and induce a chain of new defects
(see Fig.~\ref{fig:channels_open}(e)). This burst of defects will be diminished due to their
annihilation when traveling to the channel boundary, with few or none remained. After then
the similar nucleation-propagation-annihilation process repeats, resulting in the periodic
variation of defect number density as shown in Fig.~\ref{fig:channels_open}(a).

For the confined cavities studied in the last section, the enclosure constraint without 
open ends, competing with self-propelled alignment and domain flow in the bulk, 
leads to a high degree of local pattern distortion which enables the defect nucleation 
at boundary even for no-cusp, circular cavity of large enough aspect 
ratio between the lateral cavity dimension and the smectic layer spacing ($\gtrsim 25$ 
as estimated from simulations; see Fig.~\ref{fig:transitions}). This is different 
from the above results for open channels (and the results below for rings or 
annulus) where no defect nucleation would occur at smooth boundaries without cusps, 
due to different confinement topology. The mechanism originated from cusp singularity 
of confinement would play a key role on enhancing defect generation, and importantly, 
on controlling the time-correlated property of defect variations, including the 
oscillatory behavior of defect correlation for cavities with small number of cusps 
as observed in simulations. When the cusp number increases the oscillation of time 
correlation function would be damped, and thus the degree of defect periodic variation 
be reduced, as a result of the interference between the effects induced by different
individual cusps. This interference effect can account for the transition from oscillatory
to non-oscillatory decay of the correlation shown in Fig.~\ref{fig:cavities_Corr}.
In the other limit of zero cusp without the oscillatory mechanism, such as the circular
cavity, faster decay of correlation is found (Fig.~\ref{fig:transitions}(e)). These
results further indicate that the mechanism generated by cusp singularity provides an 
effective route for controlling the collective property of defect 
dynamics and correlation.

\begin{figure}
    \centering
    \includegraphics[width=0.9\textwidth]{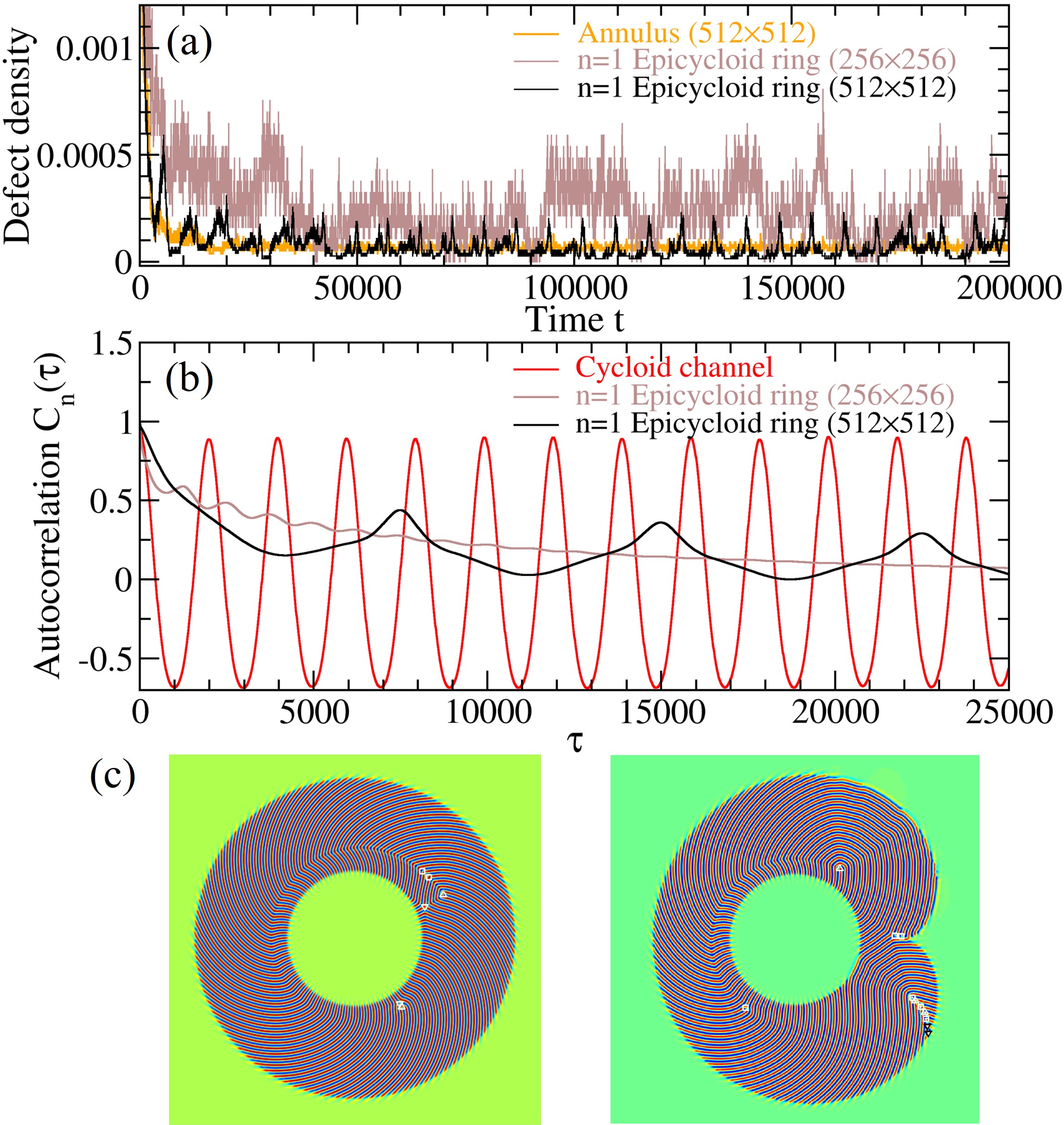}
    \caption{Defect density and time correlation for channels and rings. (a) Sample time evolution of defect number density for annulus and single-cusp epicycloid rings of two different simulation system sizes at $v_0=0.31$. (b) Autocorrelation $C_n(\tau)$ corresponding to the sample simulation of cycloid channel given in Fig.~\ref{fig:channels_open} and for $n=1$ epicycloid rings (averaged over 80 simulation runs at $t=10^5$--$10^6$ for each system size). (c) Sample snapshots of the annulus and one-cusp epicycloid ring simulated.}
    \label{fig:rings_epi1cusp}
\end{figure}

Applying this mechanism, one can expect that further confining via both inner and outer boundaries, i.e., a closed channel with a void, would result in the damping of correlation due to greater constraint and larger degree of interference between boundaries, although with similar processes of defect generation, traveling, and annihilation. This is seen in Fig.~\ref{fig:rings_epi1cusp} where the single-cusp epicycloid ring can be viewed as a curved analog of the cycloid channel given in Fig.~\ref{fig:channels_open}. For small system size (e.g., $256 \times 256$ grid points, same as Fig.~\ref{fig:channels_open}) with narrow channel width, much weaker oscillation and a damped behavior of time correlation occurs, as compared to the time-periodic behavior for the open cycloid channel (see Fig.~\ref{fig:rings_epi1cusp}(b)). Interestingly, wider channel width leads to longer period of time correlation of defect number variations, as seen from the $C_n(\tau)$ result presented in Fig.~\ref{fig:rings_epi1cusp}. (It is possible that defects could be trapped inside a wide enough epicycloid channel, without any new boundary defects nucleated, as found in roughly half of the independent simulation runs conducted for system size of $512 \times 512$ grid points. Those cases are not used in the calculation of $C_n(\tau)$ in Fig.~\ref{fig:rings_epi1cusp}.)
 In comparison, for annulus with smooth boundary, no defect multiplication occurs, analogous to the case of smooth S-shaped channel, and the bulk defects could self-circle persistently as a result of self-propulsion (see Supplementary Movie 9 for an example of annulus at $v_0=0.31$).

\begin{figure}
  \centerline{\includegraphics[width=0.9\textwidth]{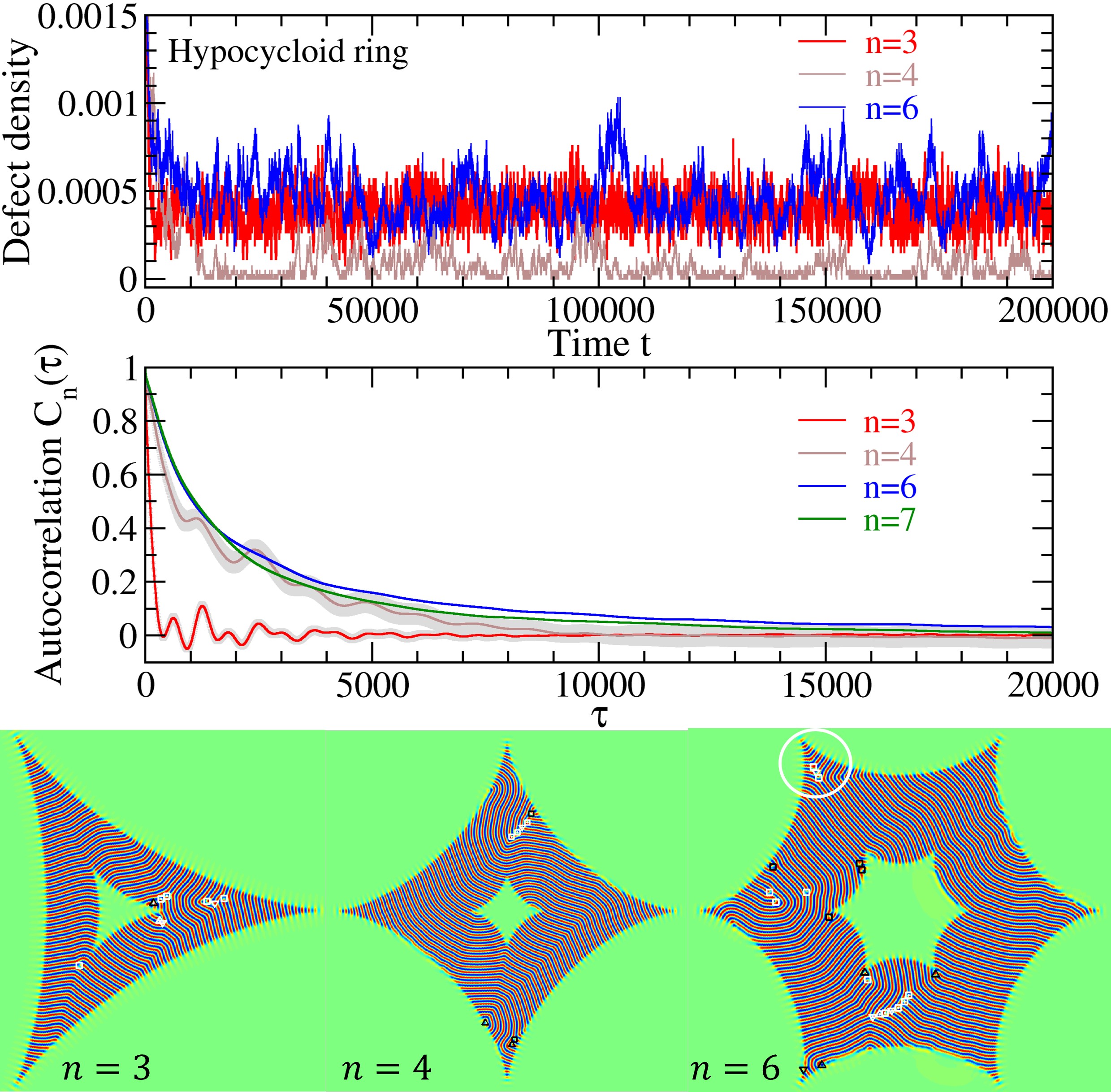}}
  \caption{Defect density and time correlation for hypocycloid rings. Sample time evolution of defect number density and the results of autocorrelation $C_n(\tau)$ (each averaged over 80 runs for $t=10^5$--$10^6$ with system grid size $512 \times 512$) for various hypocycloid rings at $v_0=0.31$. Sample snapshots of the $\psi$ field spatial profile are also shown, with the white-circled region corresponding to the local defect trapping near a cusp of $n=6$ hypocycloid ring as seen in Supplementary Movie 8.}
  \label{fig:rings_hypo}
\end{figure}

Higher degree of correlation damping is expected for closed channels/rings with more cusps interference, as verified in Fig.~\ref{fig:rings_hypo} for hypocycloid rings with $n$ varying from $3$ to $7$. Weak oscillation of $C_n(\tau)$ is found for $n=3$ and $4$; between them the $n=3$ hypercycloid ring exhibits faster decay of correlation over time, which can be attributed to its narrower channel width. Larger number of cusps (e.g., $n=6, 7$) results in non-oscillatory, damping behavior of correlation due to increasing interference between cusps, with faster damping for larger $n$ (see Fig.~\ref{fig:rings_hypo} if comparing $n=7$ to $n=6$), as expected. On the other hand, the confinement of closed channel leads to the increase of the correlation time as compared to the closed cavity, as seen from the $C_n(\tau)$ plots for $n=6$ hypocycloid cavity vs ring in Figs.~\ref{fig:cavities_Corr} and \ref{fig:rings_hypo}.

The above results for cusped channels or rings correspond to the intermediate regime of activity with $v_0$ larger than but near $v_{0c}$ showing high defect number fluctuation and defect creation-annihilation. In another dynamical regime of larger activity, the effect of self-driving dominates over that of rigid boundary confinement, leading to the absence of defect multiplication as in the case of closed cavities. In the closed channels of annular or ring geometries the defect dynamics manifests itself either via local persistent variations involving bulk grain boundaries and spirals confined by the channel walls, or interestingly, in the form of persistent self-circling current of bulk defects (similar to that found in Supplementary Movie 9 for annulus at smaller $v_0$).

We finally remark that defect pinning or local trapping can be identified under two different conditions. (i) For small self-driving strengths with $v_0<v_{0c}$, defects are pinned near the cusps as well as in the bulk (see the first column of Figs.~\ref{fig:transitions}(a)-\ref{fig:transitions}(c)), with similar results obtained for different confining geometries and topologies including closed cavities and open and closed channels as well as different cusp shapes such as the polygonal confining geometry as found in our simulations (see Supplementary Note 1 for some sample results). (ii) In the intermediate range of $v_0$ with states of highly fluctuating defect dynamics, it is possible that some defects could be temporarily trapped close to the cusp as caused by strong boundary confinement particularly in narrow closed channels, as shown in an example of an $n=6$ hypocycloid ring in Fig.~\ref{fig:rings_hypo} (see the white-circled corner) and Supplementary Movie 8. Although these look similar to that found for self-propelled rods with smectic defect structures trapped in a wedge by simulations and experiments \cite{KaiserPRL12,KumarPRE19}, the underlying mechanisms are different. In those previous studies the trapping or escape of active rods and smectic defects was for an isolated cusp/wedge in an open environment of particles. While there is an active smectic structure pinned by the open cusp, the outside region is pretty dilute in active particles and as a consequence the cusp angle and shape play a crucial role \cite{KaiserPRL12,KumarPRE19}. In contrast, here the active smectics fill the whole cavity and are overall confined. Therefore defect pinning or trapping either occurs as a result of too small activity $v_0$ to overcome the large pinning barrier in the full strongly segregated pattern, the occurrence of which is independent of cusp geometry, or at the intermediate value of $v_0$, could temporarily appear due to local strong confinement of narrow channels that hinders the propagation of boundary-nucleated defects and any subsequent defect multiplication. Finally we also mention that the classical sources of defect generation and multiplication by crystal deformation such as the Frank-Read source \cite{Frank-ReadPR50} would not directly apply here since our effects are controlled by a combination of activity (with the self-propulsion of smectic layers) and confined boundary conditions.

\section*{Conclusions}

We have examined the dynamics of topological defects in active smectic systems subjected to three types of boundary confinements, i.e., closed cavities, open and closed channels with various geometries. Our simulations based on active PFC modeling indicate a viable way to effectively vary or control the complex dynamics and collective time-variation properties of defects through both particle self-driving and the geometry and topology of strong confinement, with the underlying mechanisms intrinsically distinct from those of passive patterns and active nematic systems. These confined nonequilibrium smectic systems are featured by three distinct regimes of active and persistent defect dynamics, including defect pinning in a glassy state with ultraslow evolution, the fast self-rotating of spiral defects in cavities or local defect variations or persistent self-circling defect currents in closed channels, and interestingly, a dynamical state governed by far-from-equilibrium, nonrelaxational processes with large defect fluctuations in the transition between localized and traveling smectic patterns. For the latter, a key factor is the intermittent but perpetual creation of new defects as enabled by the confinement boundary and enhanced by cuspate boundary geometry. A transition from random to time-periodic process of defects creation and annihilation can be made possible through the control of boundary cusp singularity as the mechanism of confinement-induced defect generation. These predictions can be examined and achieved in experiments on e.g., dense self-propelled rods \cite{REVbaerSPR2020} which form an active smectic phase. Examples range from vibrated granular rods which can be exposed to circular \cite{narayan2006nonequilibrium}, epicycloid-like flower-shaped \cite{DeseignePRL10} or annular \cite{armas2020} confinements, to active colloidal rods \cite{VutukuriSoftMatt16,WykesSoftMatt16} in channels and cavities.

\section*{Methods}

\subsection*{Confinement geometries}
As described above in Eqs. (\ref{eq:F_anch}) and (\ref{eq:Vb}), to implement the planar 
anchoring condition of boundary confinement we develop an approach by combining a formulation
of surface/interface free energy for imposing the rigid boundary conditions
(i.e., $\psi=\psi_b$, $\hat{\mathbf n} \cdot \bm{\nabla} \psi = 0$, and
$\mathbf{P}=\mathbf{P}_b$, with $\psi_b$ and $\mathbf{P}_b$ the boundary values
of density field $\psi$ and orientation field $\mathbf{P}$), with the control of
confinement geometry via the spatially dependent interface energy amplitude
$V_b (\mathbf{r}) = V_{b0} \phi(\mathbf{r})$, where $V_{b0}$ is the anchoring strength.
Here $\phi(\mathbf{r})$ is an auxiliary phase field function used in the diffuse domain
method \cite{LiCMS09} and is approximated as
\begin{equation}
  \phi(\mathbf{r}) = \frac{1}{2}
  \left [ 1 + \tanh \left ( \frac{r_s(\mathbf{r})}{\Delta} \right ) \right ],
  \label{eq:phi}
\end{equation}
where $\Delta$ is the interface width of the boundary and $r_s(\mathbf{r})$ is the signed
distance function from any location $\mathbf{r}$ to the domain boundary $\partial\Omega$.
$r_s(\mathbf{r})=-d(\mathbf{r})<0$ inside the domain $\Omega$ and $r_s(\mathbf{r})=d(\mathbf{r})>0$
outside $\Omega$, with $d(\mathbf{r})$ the distance function to $\partial\Omega$.
Various methods or algorithms for calculating signed distance functions have been available.
For 2D domain boundaries studied here, we can directly obtain analytic forms of $r_s$ for some
simple geometries (see below). In the cases of complex boundary geometries, a straightforward
way of approximation, as used in our simulations, is to numerically compute the distances
$d_i = \vert \mathbf{r} - \mathbf{r}_{bi} \vert$ to the points $\mathbf{r}_{bi} \in \partial\Omega$
($i=1,2,...$) discretized on the boundary curve and find the shortest distance among them, i.e.,
\begin{equation}
  d(\mathbf{r}) = \min (d_i(\mathbf{r}))
  = \min\limits_{\mathbf{r}_{bi} \in \partial\Omega} (\vert \mathbf{r} - \mathbf{r}_{bi} \vert),
\end{equation}
which also determines the boundary point $\mathbf{r}_b$ corresponding to each $\mathbf{r}$.
The local unit normal to the boundary is then
\begin{equation}
  \hat{\mathbf n} = (\mathbf{r} - \mathbf{r}_b) / d(\mathbf{r}).
\end{equation}

The following types of confinement geometries have been examined in our simulations:

(i) \textit{Circular cavity and annulus}: For a 2D circular cavity of radius $r_0$, with cavity
center located at $\mathbf{r}_c=(x_c,y_c)$, we have $\vert\mathbf{r}\vert=r=[(x-x_c)^2+(y-y_c)^2]^{1/2}$,
$\hat{\mathbf n}=\hat{\mathbf r}$, $r_s = r-r_0$, and
\begin{equation}
  \phi(\mathbf{r}) = \frac{1}{2}
  \left [ 1 + \tanh \left ( \frac{r-r_0}{\Delta} \right ) \right ].
\end{equation}
For an annulus with inner and outer radius of $r_{\rm in}$ and $r_{\rm out}$ respectively,
$\hat{\mathbf n}=\hat{\mathbf r}$ and
\begin{equation}
  \phi(\mathbf{r}) = \frac{1}{2}
  \left \{ 1 + \tanh \left [ \frac{\left \vert r-(r_{\rm in}+r_{\rm out})/2 \right \vert -(r_{\rm out}-r_{\rm in})/2}
    {\Delta} \right ] \right \}.
\end{equation}

(ii) \textit{S-shaped open channel}: Assume the channel is aligned vertically, with its center
at $\mathbf{r}_c=(x_c,y_c)$ and the average locations of right and left boundaries at $x-x_c=\pm x_0$.
The boundary curves are of the form
\begin{equation}
  x_b = x_c \pm x_0 - S_0 \sin [q_s(y-y_c)],
\end{equation}
where $S_0$ is the amplitude and $2\pi/q_s$ is the periodicity of the S-shaped modulation.
The boundary normal is given by
\begin{equation}
  \hat{\mathbf n} = \frac{\left ( -1, dx/dy \right )}{\sqrt{1 + \left ( dx/dy \right )^2}}
  = \frac{\left ( -1, -q_s S_0 \cos [q_s(y-y_c)] \right )}{\sqrt{1 + q_s^2 S_0^2 \cos^2 [q_s(y-y_c)]}}.
\end{equation}
The area of this open channel in a system of vertical length $l_y$ is equal to $2x_0l_y$ (when $l_y$
is set as an integer number of modulation periodicity), which is used in the calculation of defect
number density. The corresponding phase field function is approximated by
\begin{equation}
  \phi(\mathbf{r}) \simeq \frac{1}{2}
  \left [ 1 + \tanh \left ( \frac{\vert x-x_c \vert - \vert x_b \vert}{\Delta} \right ) \right ].
\end{equation}
Note that although $r_s \neq \vert x-x_c \vert - \vert x_b \vert$ for this S-shaped channel, the above equation is still
a good approximation for $\phi(\mathbf{r})$ when $\Delta$ is small enough (i.e., for sharp boundary
interface).

(iii) \textit{Epicycloid cavities with $n$ cusps}: The corresponding parametric equations are
given by
\begin{eqnarray}
  && x = (a+b) \cos\theta - b \cos \left (\frac{a+b}{b}\theta \right ), \nonumber\\
  && y = (a+b) \sin\theta - b \sin \left (\frac{a+b}{b}\theta \right ), \label{eq:epicycloid}
\end{eqnarray}
where the parameter $\theta$ (not the polar angle) ranges from 0 to $2\pi$, and $b=a/n$ for
a $n$-cusped epicycloid. The area of the enclosed cavity is $(n+1)(n+2)\pi b^2$ for integer $n$.
The related phase field function $\phi(\mathbf{r})$ is calculated via Eq.~(\ref{eq:phi}) with
$r_s(\mathbf{r})$ and the unit normal $\hat{\mathbf n}$ identified numerically as described above.
Some sample results obtained from our simulations are presented in Figs.~\ref{fig:transitions}
and \ref{fig:cavities_Corr} and in
Supplementary Movies 1-4 for different cusp number $n$, including $n=3$ (of shape similar to trefoil),
$4$ (similar to quatrefoil), $5$ (ranunculoid), and $6$.

(iv) \textit{Hypocycloid cavities and rings}: For a hypocycloid cavity with $n$ cusps, the
parametric equations are written as
\begin{eqnarray}
  && x = (a-b) \cos\theta + b \cos \left (\frac{a-b}{b}\theta \right ), \nonumber\\
  && y = (a-b) \sin\theta - b \sin \left (\frac{a-b}{b}\theta \right ), \label{eq:hypocycloid}
\end{eqnarray}
where $b=a/n$. The cavity area is equal to $(n-1)(n-2)\pi b^2$. At each position inside or
outside of cavity, values of phase field function $\phi$, $r_s$, and $\hat{\mathbf n}$ are
computed numerically by following the above procedure. Some sample simulation results for
$n=4$ (astroid), $5$, and $6$ hypocycloid cavities are given in Figs.~\ref{fig:transitions}
and \ref{fig:cavities_Corr} and in Supplementary Movies 5-7. Similar setup can be used for 
hypocycloid rings, with inner and outer boundary curves each determined by the above 
parametric equations with two sets of $a$ and $b$ parameters (see some sample simulation 
snapshots given in Fig.~\ref{fig:rings_hypo} and Supplementary Movie 8).

(v) \textit{Cycloid open channel}: For a vertically aligned channel, the left boundary is
a straight line located at $x=-x_1$ (so that $r_s=\pm \vert x+x_1 \vert$ and $\hat{\mathbf n}=(1,0)$),
while the right boundary curve is of a cycloid or trochoid form described by
\begin{eqnarray}
  && x = a - b\cos\theta + x_0, \nonumber\\
  && y = a\theta -b\sin\theta.
\end{eqnarray}
It is a curtate cycloid if $a>b$, a prolate cycloid if $a<b$, and a cycloid when $a=b$ which
is used in our simulations. We choose $-\pi \leq \theta \leq \pi$ and set $2\pi a=l_y$ to satisfy
the periodic boundary condition along the $y$ direction with open ends of the channel. The
corresponding channel area is equal to $\pi(2a^2+b^2)+2\pi a(x_0+x_1)$. This channel configuration
is implemented in our simulations through numerical calculations of $\phi$, $r_s$, and
$\hat{\mathbf n}$, with examples given in Fig.~\ref{fig:channels_open}.

In principle any other types of boundary geometries, as long as the corresponding analytic or
numerical expressions of boundary curves are available, can be described via similar procedure
and thus implemented in our modeling and simulations. This approach that we introduce here,
based on Eqs. (\ref{eq:F_anch}) and (\ref{eq:Vb}) and the above implicit representation of domain boundary,
allows us to apply the pseudospectral method with periodic boundary conditions in the whole system
to numerically solve the active PFC equations subjected to the confinement of various types of
cavity or channel geometry.

\subsection*{Algorithm for defect detection}
To identify the topological defects (dislocations, disclinations, and grain boundaries) in the
simulated smectic pattern, we use an algorithm based on the combination of two methods given
in Refs.~\cite{QianPRE03,HarrisonThesis99}, with some modifications and extension. The
implementation steps are described below.

Given the local stripe orientation $\hat{\mathbf n}_s = \bm{\nabla}\psi / \vert \bm{\nabla}\psi \vert =
(\cos\theta_s, \sin\theta_s)$ with $\theta_s$ the local orientation angle of the smectic layer, 
we can calculate at each spatial location $\mathbf{r}=(x,y)$
\begin{equation}
  \vert\bm{\nabla}\psi\vert^2 \sin 2\theta_s = 2 (\partial_x\psi) (\partial_y\psi), \qquad
  \vert\bm{\nabla}\psi\vert^2 \cos 2\theta_s = (\partial_x\psi)^2 - (\partial_y\psi)^2.
\end{equation}
Then a Gaussian smoothing of each of $\vert\bm{\nabla}\psi\vert^2 \sin 2\theta_s$ and
$\vert\bm{\nabla}\psi\vert^2 \cos 2\theta_s$ is conducted over a neighboring square range of grid points
for each position $\mathbf{r}$ \cite{HarrisonThesis99}, and the local director orientation is
identified by
\begin{equation}
  \theta_s = \frac{1}{2} \arctan \left [ \frac{\left ( \vert\bm{\nabla}\psi\vert^2 \sin 2\theta_s
  \right )_{\rm smoothed}} {\left ( \vert\bm{\nabla}\psi\vert^2 \cos 2\theta_s \right )_{\rm smoothed}} \right ].
\end{equation}

To detect the locations of defect cores, at each grid point the local orientation gradient is
calculated \cite{QianPRE03}, i.e.,
%\begin{equation}
$A_s = \vert \bm{\nabla} \theta_s \vert^2$.
%\end{equation}
If $A_s$ exceeds a threshold value $A_{0s}$ (e.g., $A_{0s}=0.2/(\Delta x)^2$, with $\Delta x = \pi/4$
the numerical grid spacing), the corresponding grid point is considered to be in a defect core region.
To obtain the specific location of each individual defect core, first the individual cluster of sites
for each defect core region is identified by using the Hoshen-Kopelman (Union-Find) algorithm with
raster scan to connect neighboring grid points of each cluster tree with large enough local orientation
variation ($A_s > A_{0s}$). The cluster's center of mass then gives the position $\mathbf{r}_{\rm CM}$
of the corresponding defect core, with
\begin{equation}
  \mathbf{r}_{\rm CM} = \frac{\sum_j \mathbf{r}_j A_s(\mathbf{r}_j)}{\sum_j A_s(\mathbf{r}_j)},
\end{equation}
where $\mathbf{r}_j$ is the spatial coordinate of site $j$ within the cluster.

To reduce the artifacts or ambiguities caused by the choice of threshold $A_{0s}$, if the size of a
cluster is larger than a limit (e.g., 20 grid sites) this part is then re-clustered through the
Union-Find algorithm to divide it into smaller sub-clusters by increasing its threshold value $A_{0s}$
by a percentage (e.g., $1/8$) of $\max(A_s)-\min(A_s)$ of that cluster. In addition, if the distance
between the centers of mass $\mathbf{r}_{\rm CM}$ of any two clusters is less than another threshold value
(e.g., $5.5 \Delta x$), they will be merged if the merged/connected cluster size would not exceed an upper
limit (e.g., 18 sites). This reclustering-merging process is conducted only once, and the corresponding
defect core locations (i.e., cluster centers of mass) will be recalculated.

To identify the specific type of each individual defect, we follow the standard procedure of calculating the
topological charge (winding number) of each defect core by performing a closed-path integral of $\theta_s$
over a counterclockwise square loop around the position of defect core \cite{QianPRE03,HarrisonThesis99}.
The defect type (charge-0 dislocation vs $\pm 1/2$ disclination) is determined via the calculated 
value of topological charge. It is noted that all the above calculations are for the orientation
of stripes (determined by the apolar density field $\psi$) and the corresponding topological charges, but not 
for the polar vector field $\mathbf{P}$ which would yield different topological charges via a similar procedure 
of calculation. A boundary defect is labeled if the location of its defect core is within a certain distance
(e.g., 8 grid points) to the cavity or channel boundary. We can also identify the defect cores (clusters)
belonging to a grain boundary (cluster chain), via the Union-Find algorithm again (but not merging them),
if the distance between the centers of mass ($\mathbf{r}_{\rm CM}$) of any two clusters (defect cores) is
less than or equal to a value (e.g., $25 \Delta x$) and if there are at least $N_{\rm GB}$ (e.g., $=4$) of
such clusters (cores).

There would still be some ambiguities/uncertainties of defect identification, which are unavoidable for
any detection algorithm particularly for the cases of close or crowded defect cores. We have checked
the results by varying different parameters of the algorithm and comparing with some manual spot checks
to identify the close-to-optimal or compromised choices of parameters, and to ensure the results are
consistent statistically.

\section*{Data availability}
The data that support the findings of this study, including those for the plots of defect number density, time autocorrelations, and power spectrum, are available in Supplementary Data. All other data are available from the corresponding author upon reasonable request.

\section*{Code availability}
Some codes that support the findings of this study are available from the corresponding author upon reasonable request.

\bibliography{bibliog}

\section*{Acknowledgments}
Z.-F.H. acknowledges support from the National Science Foundation under Grant No.
DMR-2006446. H.L. and A.V. were supported by the German Research Foundation (DFG) 
via projects LO 418/20-2 and VO 899/19-2.

\section*{Author contributions}
Z.-F.H. implemented the codes and performed the simulations. Z.-F.H., H.L., and A.V. developed the theory, analysed the data, and wrote the manuscript. 

\section*{Competing interests}
The authors declare no competing interests.

\end{document}

\title{Supplementary Information \\
  Defect dynamics in active smectics induced by confining geometry and topology}
\author{Zhi-Feng Huang}
\affiliation{Department of Physics and Astronomy, Wayne State University, Detroit, MI 48201, USA}
\author{Hartmut L\"owen}
\affiliation{Institut f\"ur Theoretische Physik II: Soft Matter, Heinrich-Heine-Universit\"at D\"usseldorf,
  40225 D\"usseldorf, Germany}
\author{Axel Voigt}
\affiliation{Institute of Scientific Computing, Technische Universit\"at Dresden, 01062 Dresden, Germany}
\maketitle

\renewcommand{\theequation}{S\arabic{equation}}
\renewcommand{\thefigure}{S\arabic{figure}}
\renewcommand{\bibnumfmt}[1]{[S#1]}
\renewcommand{\citenumfont}[1]{S#1}

\section{Supplementary Note 1: Polygonal cavities and rings}

\begin{figure}[htb]
  \centering
  \includegraphics[width=0.9\textwidth]{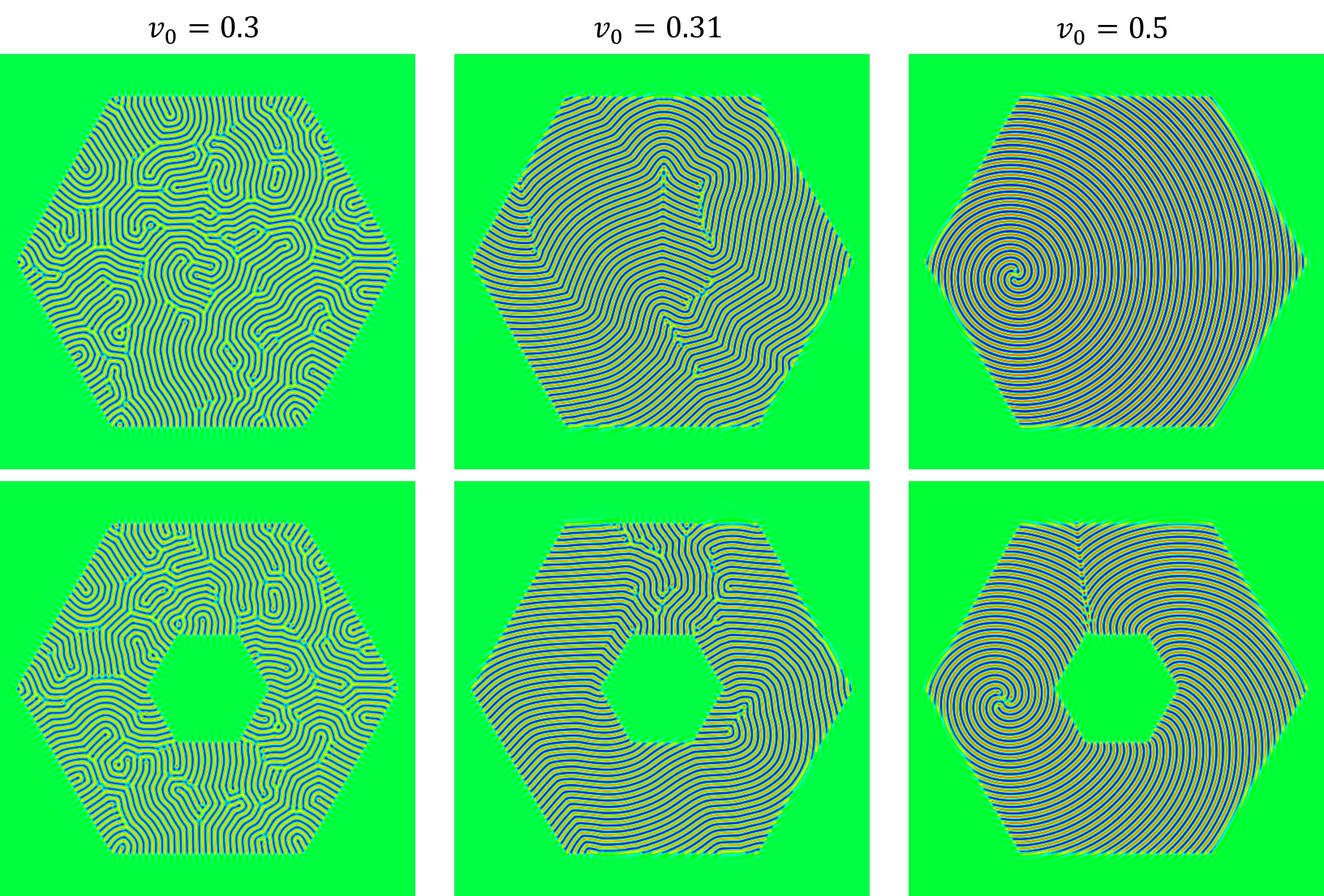}
  \caption{Sample simulation snapshots of the spatial profile of density field $\psi$ for closed
    cavities and rings with hexagonal geometry at time $t=2\times 10^5$, for $v_0 =0.3$, $0.31$,
    and $0.5$ corresponding to three different dynamical regimes of defects.
  }
  \label{fig:hexagon_cr}
\end{figure}

We have performed additional simulations for active smectics confined in cavities and rings with
polygonal geometries. The results obtained are similar to those of epicycloid and hypocycloid
geometries reported in the main text, including the three dynamical regimes characterized by
states of defect pinning, highly fluctuating defect dynamics (creation-propagation-annihilation),
and defect self-rotation or local variations, with the increase of self-propulsion strength $v_0$.
Some sample simulation results for hexagonal confining geometry with two different topologies,
i.e., closed cavities and rings, are given in Fig.~\ref{fig:hexagon_cr}.

